\documentclass[a4paper,amsmath,amssymb,superscriptaddress,twocolumn,showkeys,showpacs,prl,aps, 10pt]{revtex4-1}
\usepackage{color}
\usepackage{graphicx}
\usepackage{hyperref}

\newcommand{\dagga}{{\phantom{\dagger}}}
\newcommand{\bk}{\mathbf{k}}
\newcommand{\be}{\begin{equation}}
\newcommand{\ee}{\end{equation}}
\newcommand{\bea}{\begin{eqnarray}}
\newcommand{\eea}{\end{eqnarray}}
\newcommand{\ba}{\begin{eqnarray*}}
\newcommand{\ea}{\end{eqnarray*}}
\newcommand{\up}{\uparrow}
\newcommand{\down}{\downarrow}
\newcommand{\eqn}[1]{(\ref{#1})}
\newcommand{\fract}[2]{\frac{\displaystyle #1}{\displaystyle #2}}

\begin{document}

\title{Ferromagnetic Kondo effect in a triple quantum dot system}

\author{P. P. Baruselli}
\affiliation{SISSA, Via Bonomea 265, Trieste 34136, Italy}
\affiliation{CNR-IOM, Democritos Unit\'a di Trieste, Via Bonomea 265, Trieste 34136, Italy}

\author{R. Requist}
\affiliation{SISSA, Via Bonomea 265, Trieste 34136, Italy}
\affiliation{CNR-IOM, Democritos Unit\'a di Trieste, Via Bonomea 265, Trieste 34136, Italy}

\author{M. Fabrizio}
\affiliation{SISSA, Via Bonomea 265, Trieste 34136, Italy}
\affiliation{CNR-IOM, Democritos Unit\'a di Trieste, Via Bonomea 265, Trieste 34136, Italy}

\author{E. Tosatti}
\affiliation{SISSA, Via Bonomea 265, Trieste 34136, Italy}
\affiliation{CNR-IOM, Democritos Unit\'a di Trieste, Via Bonomea 265, Trieste 34136, Italy}
\affiliation{ICTP, Strada Costiera 11, Trieste 34014, Italy}

\date{\today}

\begin{abstract}
A simple device of three laterally-coupled quantum dots, 
the central one contacted by metal leads, 
provides a realization of  the ferromagnetic Kondo model, which is characterized by interesting properties like a  non-analytic 
inverted zero-bias anomaly and an extreme sensitivity to a magnetic field. Tuning the gate voltages 
of the lateral dots allows to study the transition from ferromagnetic to antiferromagnetic Kondo effect, 
a simple case of a Berezinskii-Kosterlitz-Thouless transition. We model the device by three coupled Anderson impurities 
that we study by numerical renormalization group. We calculate the single-particle spectral function of the central dot, 
which at zero frequency is proportional to the zero-bias conductance, across the transition, both in the absence and 
in the presence of a magnetic field. 
\end{abstract}

\maketitle


In spite of its simplicity -- being just a magnetic impurity embedded in a conduction bath -- the Kondo model 
exhibits rich many-body
physics\cite{Hewson} that continues to attract scientific interest in a variety of contexts. 
It is common to distinguish between Kondo models with Fermi-liquid properties 
{\it a' la} Nozi\'eres\cite{Nozieres-JLTP} and 
those that instead display non-analytic, hence non-Fermi-liquid, behavior as a function of state variables. 
The latter class includes under- and over-screened Kondo models\cite{nozieres,Affleck&Ludwig,Hewson} 
as well as clusters of 
magnetic impurities in particular circumstances\cite{Affleck&Ludwig&Jones,Affleck-trimer,Ferrero&DeLeo}. 
Non-Fermi liquid properties are not common in traditional magnetic alloys,\cite{Cox&Fred} 
where the metal hosts generally
possess enough scattering channels that can
perfectly screen the magnetic impurity. They may instead
be realized in confined scattering geometries such as
a quantum dot or a magnetic atom/molecule contacted by
metal leads. Indeed, by means of such devices there are 
already many experimental realizations  of exotic non-Fermi-liquid Kondo models, 
see e.g. Refs. \cite{Parks11062010}, \cite{potok_overscreened}, \cite{Chang&Chen} and 
\cite{Crommie-trimer}. 

One case, however, which so far remains elusive 
is the ferromagnetic Kondo model (FKM)\cite{Hewson-underscreened} -- 
where
the impurity and  the conduction electrons are coupled ferromagnetically -- except for 
its indirect manifestation in the under-screened Kondo effect.\cite{nozieres,Parks11062010}
It has been proposed that the so-called giant moments induced by $3d$ transition metal impurities 
diluted in $4d$ transition metals may actually be a manifestation of FKM,\cite{paola_gentile} 
but experiments that could pin it down are still lacking.  
This is unfortunate since the FKM is the simplest example of non-Fermi liquid behavior; 
at low temperature the impurity spin behaves essentially as a free local moment apart from 
logarithmic singularities.\cite{Hewson-underscreened,Zarand-ferro}  

Here we present a possible realization of a FKM by means of three laterally-coupled quantum dots. We 
also discuss the appealing possibility of crossing the Berezinskii-Kosterlitz-Thouless (BKT) phase transition 
from ferromagnetic to antiferromagnetic Kondo effect, whose spectral weight anomaly change we study 
here by means of numerical renormalization group 
(NRG)\cite{wilson, Krishnamurthy-1,Krishnamurthy-2} in a toy-model for the device. 

\begin{figure}[ptb]
\includegraphics[width=0.3\textwidth]{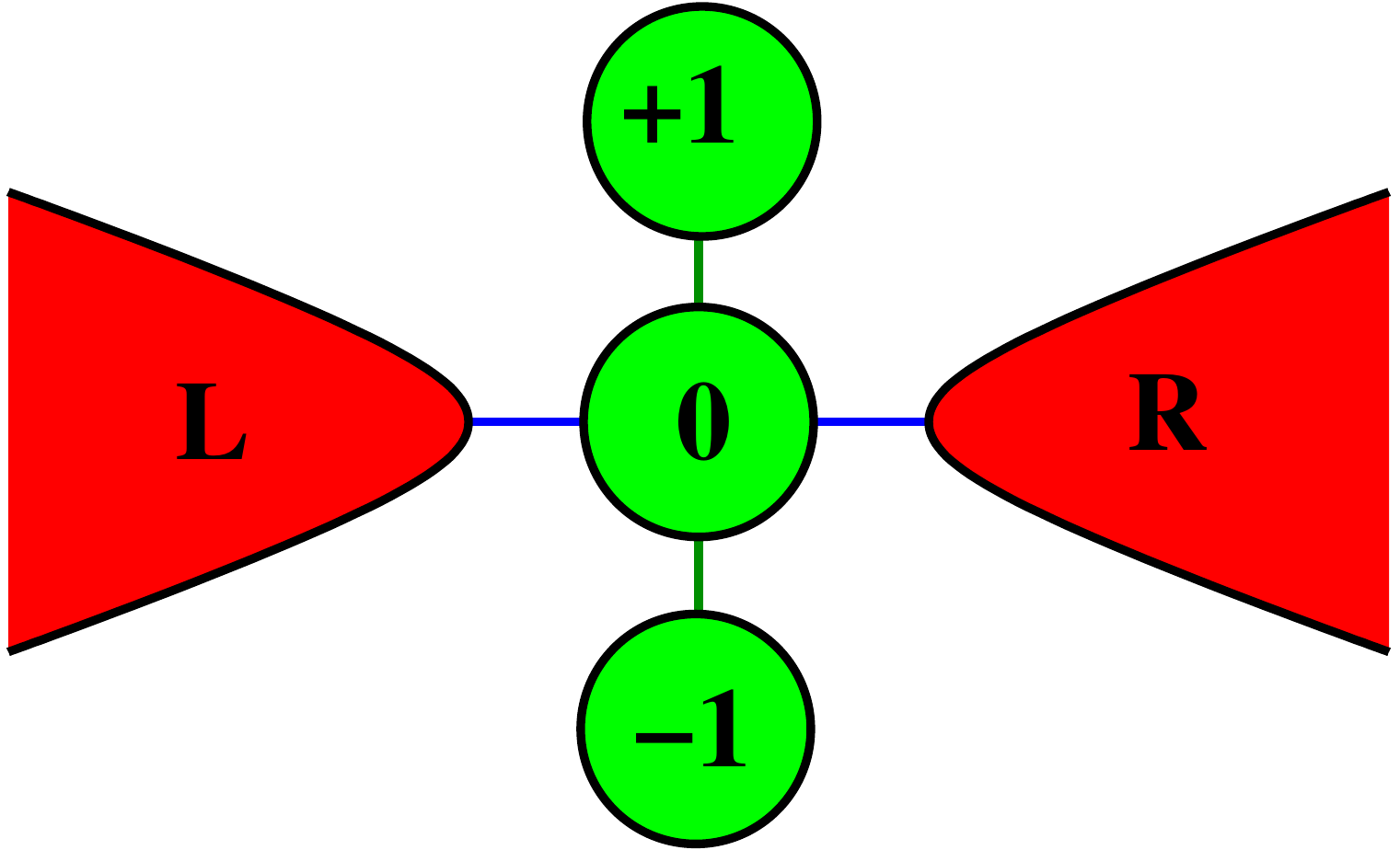}
 \caption{(Color online) A schematic representation of our device described by Eq. \ref{H3}, with three 
 quantum dots (in green). Only the central one, labelled as 0, is attached to metallic leads (in red).
} \label{fig_model}
\end{figure}

Our {\it gedanken} (but entirely feasible)\cite{triple-dots-1,triple-dots,Laird} 
set-up, schematically shown in Fig. \ref{fig_model}, consists of three quantum dots, 
labelled as $\pm 1$ and 0 in the figure, with the central one contacted by two metal leads, $R$ and $L$. 
We model the isolated three-dot device with a Hamiltonian
 \bea\label{H3}
\mathcal{H}&=&\sum_{i=-1}^{+1} \bigg(\epsilon_i n_i +\frac{U_i}{2}\big(n_i-1\big)^2 \bigg) 
\nonumber\\
&& -\sum_\sigma \big(t_-\,c^\dagger_{-1\sigma}c^\dagga_{0\sigma}+ 
t_+\, c^\dagger_{0\sigma}c^\dagga_{+1}+H.c.
\big),
\eea
where we keep just one orbital per dot, $U_i$ are the charging energies, and the dots 
are mutually coupled by single-particle tunneling. As usual, we shall assume non-interacting leads, coupled 
to the central dot 0 by tunneling. For convenience, we also assume 
equivalent $R$ and $L$ leads , so that only their symmetric combination matters in the linear response 
regime of interest to us.  Particle-hole symmetry is also assumed. The lead-dot tunneling is thus parametrized by a single quantity, the hybridization width 
$\Gamma = \pi\sum_\bk\, \left|V_\bk\right|^2\,\delta(\epsilon_\bk)$, 
where $\epsilon_\bk$ 
is the energy and $V_\bk$ the tunneling amplitude into the dot of the symmetric  $L+R$
combination of the lead electrons at momentum $\bk$.  

We further assume that each dot is brought by gate voltage into the Coulomb blockade regime with a single unpaired electron, i.e. $\epsilon_i\simeq 
0$ and $t_\pm \ll U_i$, $\forall i$.  In this limit, the isolated trimer behaves like a three-site Heisenberg model described by an effective Hamiltonian 
\be
\mathcal{H}_\text{eff} = J_+\,\mathbf{S}_0\cdot\mathbf{S}_{+1} + J_-\,\mathbf{S}_0\cdot\mathbf{S}_{-1},
\label{H-spin}
\ee  
with positive $J_\pm$, where $\mathbf{S}_i$, $i=0,\pm 1$, are spin-1/2 operators residing 
on the corresponding dots. The ground state of \eqn{H-spin} has total spin 1/2 and explicitly reads 
\be
\mid \text{GS},\sigma \rangle = \cos\theta\,\mid \text{O},\sigma\rangle - 
\sin\theta\,\mid\text{E},\sigma\rangle,\label{GS}
\ee
where $\sigma=\uparrow,\downarrow$ is the $z$-component of the total spin, and 
$\tan 2\theta = \sqrt{3}\,\big(J_+-J_-\big)/\big(J_+ + J_-\big)$. In Eq. \eqn{GS}, 
$\mid\text{O},\sigma\rangle$ is the state, odd by inversion through dot 0, obtained 
by coupling dots $+1$ and $-1$ into a triplet that is coupled to dot 0 to form a spin 1/2. 
Vice versa,  $\mid\text{E},\sigma\rangle$ is the state, even by reflection, obtained 
by coupling dots $+1$ and $-1$ into a singlet, leaving a free spin-1/2 on dot 0. 

If on dot 0 the electron is removed or one more electron is added 
through the leads, the trimer ends up in a triplet 
or singlet configuration, with probability proportional to $\cos^2\theta$ and $\sin^2\theta$, respectively. 
Specifically, if $V=\sqrt{\sum_\bk \left|V_\bk\right|^2}\ll U_0$, the Kondo exchange, $J_\text{eff}$, can be found by second order perturbation theory:
\bea
\fract{J_\text{eff}}{2V^2} &=&  
\langle \text{GS},\uparrow\mid c^\dagger_{0\uparrow}\,
\fract{1}{\mathcal{H}-E_\text{GS}}
\, c^\dagga_{0\down}\mid\text{GS},\down\rangle
\nonumber\\
&& - \langle \text{GS},\uparrow\mid c^\dagga_{0\downarrow}\,
\fract{1}{\mathcal{H}-E_\text{GS}}
\,c^\dagger_{0\up}\mid\text{GS},\down\rangle,\label{J_eff-exp}
\eea
where 
\ba
c^\dagga_{0\down(\up)}\mid\text{GS},\down(\up)\rangle &=& 
\mp \fract{\cos\theta}{\sqrt{3}}\,\mid t,0\rangle -  \sin\theta\,\mid s\rangle,\\
c^\dagger_{0\up(\down)}\mid\text{GS},\down(\up)\rangle &=& 
- \fract{\cos\theta}{\sqrt{3}}\,c^\dagger_{0\up}c^\dagger_{0\down}\mid t,0\rangle \mp   \sin\theta\,c^\dagger_{0\up}c^\dagger_{0\down}\mid s\rangle,
\ea
and where
\bea
\mid t,S_z=0\rangle &=& \frac{1}{\sqrt{2}}\big(c^\dagger_{+1\up}c^\dagger_{-1\down}
+ c^\dagger_{+1\down}c^\dagger_{-1\up}\big)\mid 0\rangle,\label{triplet}\\
\mid s\rangle &=& \frac{1}{\sqrt{2}}\big(c^\dagger_{+1\up}c^\dagger_{-1\down}
- c^\dagger_{+1\down}c^\dagger_{-1\up}\big)\mid 0\rangle,\label{singlet}
\eea
are the $S_z=0$ component of the triplet state and the singlet state, respectively. 
It follows that 
\bea
\fract{J_\text{eff}}{2V^2} &=& -\fract{\cos^2\theta}{3}\,
\langle t,0\mid \mathcal{R}\mid t,0\rangle+ \sin^2\theta\, \langle s\mid \mathcal{R}\mid s\rangle
\nonumber \\
&\equiv& -\fract{\cos^2\theta}{3}\,\gamma_t 
+\sin^2\theta\,\gamma_s,\label{J_eff}
\eea
where the resolvent operator 
\[
\mathcal{R} = \fract{1}{\mathcal{H}-E_\text{GS}} 
+ c^\dagga_{0\down}c^\dagga_{0\up}\,\fract{1}{\mathcal{H}-E_\text{GS}}\,
c^\dagger_{0\up}c^\dagger_{0\down},
\]
and $\gamma_s > \gamma_t > 0$ since the intermediate singlet has lower energy than the triplet. 
The lead-dot exchange is therefore ferromagnetic 
if $\gamma_t\,\cos^2\theta > 3\,\gamma_s\,\sin^2\theta$, and
antiferromagnetic in the opposite case.
We observe that, if inversion symmetry holds, $J_+=J_-$, then $\theta=0$ hence the 
lead-dot exchange is ferromagnetic, thus providing a realization of the FKM. 
We expect that in a real device inversion symmetry is generally broken; nevertheless there 
still is a good chance for ferromagnetism to survive in a wide region (by definition 
$\cos^2\theta\geq 3\sin^2\theta$, hence just because $\gamma_s>\gamma_t$ is it possible for  
the Kondo exchange to turn antiferromagnetic). 

In conclusion, the set-up shown in Fig. \ref{fig_model} seems indeed able to realize, 
as noted earlier,\cite{Avishai-trimer, Logan-trimer} the much-sought FKM. 
Moreover, it suggests a simple way to study experimentally the transition from the FKM to the more 
conventional antiferromagnetic Kondo model, 
first described by Anderson, Yuval and Hamann~\cite{AYH},  and expected to be of the BKT 
type. 
Indeed, changing the gate voltage $\epsilon_{+1}$ with respect 
to $\epsilon_{-1}$ drives the system further away from the inversion symmetric point,  
eventually turning the exchange from ferromagnetic to antiferromagnetic, as we are going to show in what follows. Alternatively, when inversion 
symmetry is retained, one could still drive a transition by including  a direct hopping or spin-exchange between dots +1 and -1.\cite{2stage-1,Logan-trimer} In this case, however, the transition looks profoundly  different from what we shall discuss,  
as it either reflects the level crossing between the two states 
$\mid \text{O},\sigma\rangle$ and $\mid \text{E},\sigma\rangle$,\cite{Logan-trimer} see Eq. \eqn{GS}, or, when the charging energy of dot 0 is suppressed,  
the singlet-triplet crossing\cite{2stage-1} of the above mentioned two-electron states $\mid t,S_z=0,\pm 1\rangle$ and $\mid s\rangle$,  see Eqs. \eqn{triplet} and \eqn{singlet}. A different possibility 
have been put forth by the authors of Ref. \cite{Avishai-trimer}, who argue that a transition from 
ferromagnetic to antiferromagnetic Kondo effect may occur without breaking inversion symmetry if  $U_0\ll U_{-1}=U_{+1}$ and the central dot energy $\epsilon_0$ exceeds a threshold value (see the Supplementary Material for more details).   

We shall instead give up inversion symmetry, and investigate the route to a BKT transition by tuning 
the lateral dot asymmetry $\epsilon_{+1}=-\epsilon_{-1}=\delta\epsilon$ 
with the Hamiltonian \eqn{H3} in the simple case when $t_+=t_-=t$, $U_0=U_{+1}=U_{-1}=U$, 
analyzed by means of NRG.\cite{wilson,Krishnamurthy-1,Krishnamurthy-2} 
For simplicity we take a flat conduction-band density of states, $\rho(\epsilon) = \rho_0 =1/2D$ 
when $\epsilon\in [-D,D]$ and zero otherwise, with the half-bandwidth $D$ our unit of energy. 

We have employed the ``NRG Ljubljana`` package \cite{nrg_ljubljana}, implementing the \textit{z}-averaging technique with $z=8$,\cite{zitko_pruschke_z} the full-density-matrix approach\cite{weich_dmnrg} and the self-energy trick. \cite{bulla_selfenergytrick} We used $\Lambda=2$ as the discretization parameter and a truncation cutoff of $10\,\omega_N$ ($\omega_N=\Lambda^ {-N/2}$, $N$ being the $N$-th NRG iteration). Spectral functions are computed 
by broadening delta-peaks at zero temperature with a log-Gaussian kernel \cite{bulla01} with $b=0.3$, and at finite temperature with the kernel of Ref. \cite{weich_dmnrg}.

\begin{figure}[thb]
\includegraphics[width=0.5\textwidth]{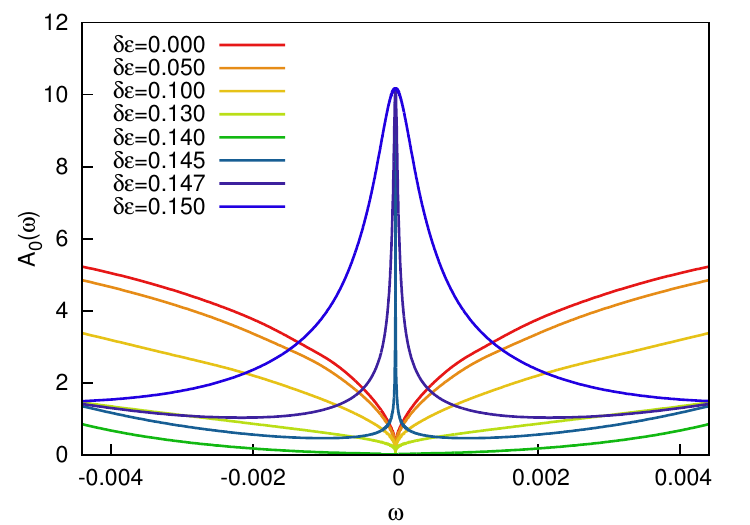}
\caption{(Color online) NRG spectral function $A_0(\omega)$ on dot 0 as a function of $\delta\epsilon$ for  
$U=0.3$, $t=0.03$, $\Gamma= 0.02\,\pi$ with a flat conduction band density of states of half-bandwidth 
$D=1$. Note the transition from the ferromagnetic Kondo, signaled by a minimum of 
$A_0(\omega)$ 
at $\omega=0$, to the regular, antiferromagnetic Kondo, where instead $A_0(\omega)$ is maximum at $\omega=0$ .}\label{DOS}
\end{figure} 

In Fig. \ref{DOS} we show the single-particle spectral density on the central dot, 
$A_0(\omega)$, 
as a function of frequency $\omega$ for different values of $\delta\epsilon$. $A_0(\omega=0)$ 
is actually proportional to the zero-bias conductance,\cite{meir} while its behavior at small $\omega$ 
mimics that of the conductance at small bias. We note that for small $\delta\epsilon$, which corresponds 
to a ferromagnetic Kondo exchange, $J_\text{eff}<0$ in Eq. \eqn{J_eff}, $A_0(\omega=0)=0$; 
the zero-bias conductance vanishes and the zero-bias anomaly is inverted. 
The spectral function displays  a logarithmic ``dimple" at $\omega =0$, associated with 
the $1/\ln^2\big(\omega/T_0\big)$ singularity typical of the FKM.\cite{Hewson-underscreened} Here 
$T_0 \propto \sqrt{\left|J_\text{eff}\right|}\,\exp{\big(-1/\rho_0\,J_\text{eff}\big)}$ is 
an energy scale that actually diverges approaching the BKT transition from the ferromagnetic side, i.e. 
$J_\text{eff}\to 0$ from below.\cite{Hewson-underscreened}

On the contrary, above a critical $\delta\epsilon_c$, 
$A_0(\omega=0)$ suddenly jumps to a finite value; the conventional 
Abrikosov-Suhl  spectral
anomaly of the regular, antiferromagnetic Kondo model is recovered. 
Because of our choice of parameters, the Hamiltonian is particle-hole symmetric hence $A_0(\omega=0)$ 
jumps from 0 to its unitary value $2/(\pi\Gamma)$ at the transition. Away from particle-hole symmetry, 
the jump still exists but will be 
smaller. The energy scale that controls the antiferromagnetic side, $J_\text{eff}>0$,  is 
of course the Kondo temperature $T_K \propto \sqrt{J_\text{eff}}\,\exp{\big(-1/\rho_0\,J_\text{eff}\big)}$, 
usually defined as the half-width at half-maximum of $A_0(\omega)$, which vanishes 
on approaching the BKT transition, $J_\text{eff}\to 0$ from above.  

We emphasize that the charging energy on the lateral dots is crucial to the existence of the 
ferromagnetic Kondo regime, which is strictly absent if $U_{+1}=U_{-1}=0$,  
as in the model of Refs. \cite{Trocha-Dicke} and \cite{Ulloa-Dicke} (see the Supplementary Material 
for more details). 

\begin{figure}[tbh]
\includegraphics[width=0.22\textwidth]{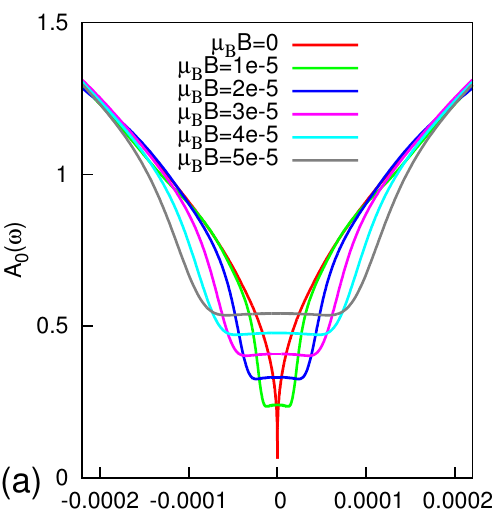}
\includegraphics[width=0.22\textwidth]{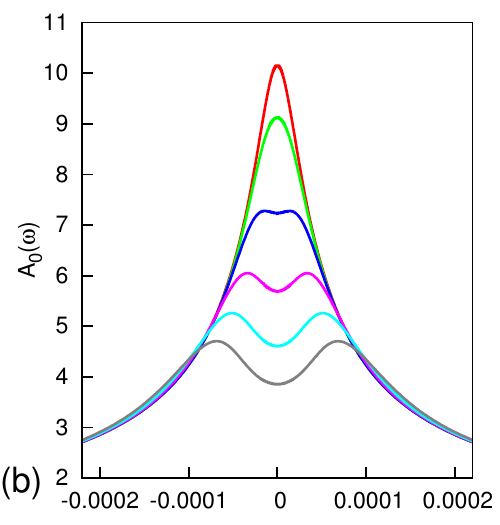}
\caption{(Color online) NRG spectral function $A_0(\omega)$ of the central dot 0, for increasing magnetic 
field $B$ (we take $\mu_B=1$ and $g=2$, hence $2B$ is the Zeeman splitting) and the same 
parameters of Fig. \ref{DOS}. Panel (a) refers to $\delta\epsilon=0$, 
while panel (b) to $\delta\epsilon=0.147$. 
In the ferromagnetic Kondo (FMK) regime, panel (a), the logarithmic dimple is immediately 
destroyed, and replaced by inelastic spin-flip excitations; in the  antiferromagnetic (standard) 
Kondo regime, panel (b), where $T_K \simeq 7\times 10^{-5}$ , 
the Abrikosov-Suhl resonance is only split by a sufficiently large field 
 $2g\mu_B B\sim k_B T_K$ .}\label{DOS-B}
\end{figure} 
\begin{figure}[tbh]
\includegraphics[width=0.22\textwidth]{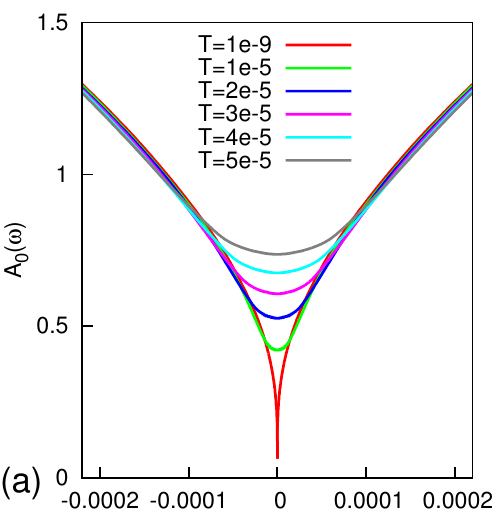}
\includegraphics[width=0.22\textwidth]{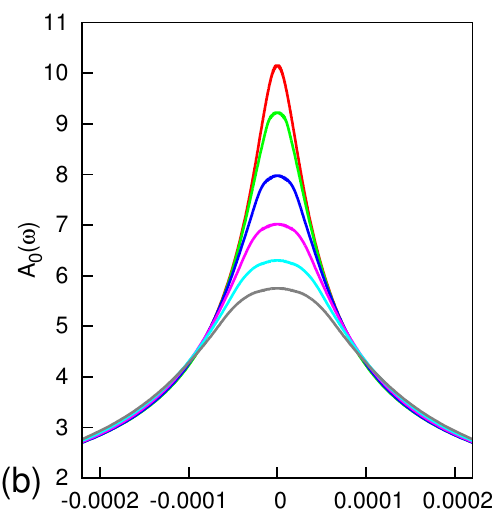}
\caption{(Color online) NRG spectral function $A_0(\omega)$ of the central dot 0, for increasing temperature $T$ and the same parameters of Fig. \ref{DOS}. Panel (a) refers to $\delta\epsilon=0$, 
while panel (b) to $\delta\epsilon=0.147$.}\label{DOS-T}
\end{figure} 

The standard, most reliable way to reveal the Kondo-like origin of a zero bias anomaly is by applying a magnetic field $B$.
In the conventional antiferromagnetic Kondo-effect, a magnetic field 
will split the  Abrikosov-Suhl resonance 
only if sufficiently large, $g \mu_B B\gtrsim 
0.5\, T_K$ \cite{Costi-magnetic-field}. This is indeed 
the case on the antiferromagnetic side of the transition, $J_\text{eff}>0$, see panel (b) of Fig. \ref{DOS-B}. 
On the contrary, 
on the ferromagnetic side, $J_\text{eff}<0$, panel (a), any magnetic field, however small, 
destroys the logarithmic dimple replacing it right away 
with a symmetric pair of inelastic spin-flip Zeeman excitations. 
In addition, $A_0(\omega=0)$,  hence 
the zero-bias conductance, increases with $B$ at low temperature, contrary to 
the antiferromagnetic side, where it drops. 
We expect moreover that a finite temperature $T$ 
will cutoff the logarithmic dimple at 
low frequency and raise up $A_0(\omega=0)\sim 1/\ln^2 (T/T_0)$, thus leading to an increase of  
zero-bias conductance, again unlike the regular antiferromagnetic Kondo effect: this is shown in Fig. \ref{DOS-T}.

\begin{figure}[tbh]
\includegraphics[width=0.4\textwidth]{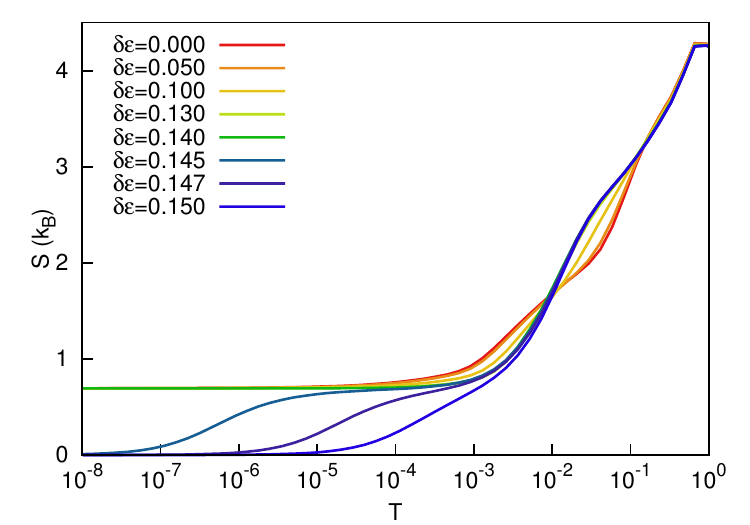}
\caption{(Color online) Entropy as a function of temperature with the same parameters as in Fig. \ref{DOS}.
} 
\label{entropy}
\end{figure}

Another quantity that transparently highlights the physics of the model is the entropy, which we plot in 
Fig. \ref{entropy} for the same values of $\delta\epsilon$ as in Fig. \ref{DOS}. We observe that, 
on the ferromagnetic side, the entropy levels off at the $\ln 2$ value of an unscreened spin-1/2 already at substantially high temperatures. On the contrary, on the antiferromagnetic side of the BKT 
transition, the entropy, after a $\ln 2$ plateau, more visible the closer the transition,  
finally drops down to zero below $T_K$.  

In conclusion, we have shown that a three dot device,
the central one contacted by metal leads, may provide a realization not only of the ferromagnetic Kondo model,\cite{Avishai-trimer,Logan-trimer} but, upon  
gating of the lateral dots, also of a Berezinskii-Kosterlitz-Thouless 
transition from ferromagnetic Kondo to regular, antiferromagnetic Kondo effect. 
The two phases should  differ sharply in their zero-bias conductance anomaly,  
the ferromagnetic one being inverted   
and very differently modified by magnetic field 
and temperature.

More generally, our proposed system illustrates a 
generic mechanism leading to FKM, namely tunneling across one orbital in presence of
other magnetic orbitals.
 This kind of situation could for example also be realized at selected surface adsorbed molecular radicals
and detected in, e.g., STS  or photoemission anomalies, an area where there is much active work. 
However, we should mention that, according to our calculations, the FKM anomalies are more 
visible when both $U$ and the tunneling amplitude into the leads are larger than the inter-dot tunneling 
$t$, which might be hard to achieve in molecular radicals.

This work was supported by PRIN/COFIN 2010LLKJBX\_004. We acknowledge useful discussions with P. Wahl and M. Ternes. We are grateful to Rok \v{Z}itko for his help with the NRG code.


%

\end{document}